\documentclass[12pt,preprint,psfig]{aastex}

\newcommand{\asca}{{\it ASCA} }

\newcommand{\xmm}{{\it XMM-Newton} }

\newcommand{\chandra}{{\it Chandra} }

\newcommand{\feka}{{Fe~K$\alpha$} }

\newcommand{\ergs}{${\rm erg \ cm^{-2} \ s^{-1}}$ }

\newcommand{\erg}{${\rm erg \ s^{-1}}$ }

\begin{document}
\title{\xmm Observations of the Seyfert 2 Galaxy NGC~7590: the Nature of X-ray Absorption}
\author{X. W. Shu\altaffilmark{1}, T. Liu\altaffilmark{1}, 
\& J. X. Wang\altaffilmark{1} }
\altaffiltext{1}{CAS Key Laboratory for Research in Galaxies and Cosmology, Department of
 Astronomy, University of Science and Technology of China, Hefei, Anhui 230026, China, xwshu@mail.ustc.edu.cn}
\begin{abstract} 
We present the analysis of three
\xmm observations of the Seyfert 2 galaxy NGC~7590. 
The source was found to have no X-ray absorption in the low spatial resolution \asca 
data. The \xmm observations provide a factor of $\sim 10$ better spatial 
resolution than previous \asca data. We find that the X-ray emission of NGC 7590 
is dominated by an off-nuclear ultra-luminous X-ray source and an extended emission from the host galaxy. The nuclear X-ray emission is rather weak comparing with the host galaxy.
Based on its very low X-ray luminosity as well as the small ratio between the 2-10 keV 
and the [O III] fluxes, we interpret NGC 7590 as Compton-thick rather than being 
an "unobscured" Seyfert 2 galaxy. 
Future higher resolution observations such as \chandra are crucial to shed light on the 
nature of the NGC 7590 nucleus.
\end{abstract}


\section{Introduction}
\label{intro}

According to the standard unification model for active galactic nuclei (AGN), 
Seyfert 2 galaxies are intrinsically the same as Seyfert 1 galaxies. 
The observed differences between the two types of Seyfert galaxies are primarily due to 
an orientation effect (Antonucci 1993). The absence of broad emission lines 
in the optical spectra of Seyfert 2 galaxies is due to obscuration 
by an optically thick structure (the so-called dusty torus) from our line of sight. 
Optical spectropolarimetry 
observations have detected hidden broad emission lines in
 a large fraction of Seyfert 2 galaxies (Moran et al. 2000; Tran 2001), providing 
strong evidence in favor of the unification model. Further support for this 
model is given by the X-ray studies that have demonstrated that many Seyfert 2 
galaxies are absorbed by large hydrogen column densities (typically 
$N_{\rm H}>10^{23}$ cm$^{-2}$; e.g., Risaliti, Maiolino \& Salvati 1999). 
    
However, recent observations have shown that a fraction of  
type 2 AGNs show no or very low X-ray absorption measured from their X-ray spectra 
($N_{\rm H}<10^{22}$ cm$^{-2}$; e.g., Pappa et al. 2001; Panessa \& Bassani 2002; 
Walter et al. 2005; Gliozzi, Sambruna \& Foschini 2007), apparently in contrast to 
the standard unification model.
Panessa \& Bassani (2002) estimate the percentage of this type of source
 in the range 10\%--20\%. 
This number is larger than that found by Risaliti et al. (1999) in a sample of 
Seyfert 2 galaxies (4\%), but consistent with the estimate made by Caccianiga et al. (2004, 12\%).
X-ray unabsorbed Seyfert 2 galaxies may result from the dust reddening intrinsic to  
the broad line region (BLR) rather than the orientation-based obscuration effects 
(Barcons, Carrera \& Ceballos 2003; Corral et al. 2005). 
Alternatively, the unabsorbed Seyfert 2 galaxies may genuinely lack the 
 BLR as a result of low accretion rate (Georgantopoulos \& Zezas 2003; Bianchi et al. 2008; 
Panessa et al. 2009). 
In other words, low-accretion rate systems may not form a stable BLR (see Nicastro 2000 for the 
model), so they would not show broad lines in their optical spectra, 
and not necessarily being obscured in X-ray.
Though the discovery of "unobscured" type 2 AGNs has important implications 
for the current unification 
model (i.e., in addition to the orientation effects, other physical effects should be invoked to 
interpret the difference observed between type 1 and type 2 AGNs), only few objects have yet been 
conclusively proven as a class.  

It has been argued that some "unobscured" Seyfert 2 galaxies are actually
 Compton-thick (e.g., Pappa et al. 2001), 
in which the direct nuclear component below 10 keV is completely suppressed and we would 
only witness an unabsorbed spectrum due to scattered nuclear radiation and/or host galaxy emission 
from a circumnuclear starburst. 
More recently, Brightman \& Nandra (2008) presented a detailed spectral 
analysis of six unabsorbed Seyfert 2 galaxies, and found that four 
out of them are in fact heavily obscured (see also Ghosh et al. 2007). 
{ In addition, Shi et al. (2010) presented a multi-wavelength study of a sample of 
unobscured Seyfert 2 galaxies, 
and found that most of them are actually intermediate-type AGNs with broad emission lines or 
Compton-thick sources.}  
Therefore, one has to be cautious 
in identifying the candidates of the unabsorbed Seyfert 2 galaxies.


In this paper, we present new \xmm observations of NGC 7590, one of the 
so-called "unobscured" Seyfert 2 galaxies.
NGC 7590 ($z=0.005$) belongs to the extended 12~$\mu$m sample (Rush, Malkan \& Spinoglio 1993).
No broad H$\alpha$ emission line is visible in either direct or polarized light (
Heisler, Lumsden \& Bailey 1997). 
The first X-ray spectrum observed by \asca was analyzed by Bassani et 
al (1999). Their best-fit model is an unabsorbed power law 
with $\Gamma=2.29^{+0.20}_{-0.13}$, and N$_H<9.2\times10^{20}$ cm$^{-2}$.
No Fe K emission line is detected in the \asca data. $ROSAT$ HRI imaging has shown that there is an 
off-nuclear ultra-luminous X-ray source (ULX) that is probably 
associated with the galaxy NGC 7590 (Colbert \& Ptak 2002). 
The better spatial resolution of \xmm allows us to see more directly the X-ray
 emission from the nuclear region. 
In the following,   
a cosmology with $H_0 = 70$ km s$^{-1}$ Mpc$^{-1}$, $\Lambda=0.73$, and $\Omega=1$ 
is assumed throughout.


\section{Observation and Data Reduction}
\label{obs}

The \xmm pointing observation of NGC 7590 (denoted OBS 1)
was made on 2007 April 30 for a duration of about 45 ks. 
Besides the pointing observation, the source was also 
detected $\sim$ 9 arcmin away from the center of the 
field of view (FOV) of NGC 7582 \xmm observations during revolution 267 
(2001 May, OBS 2) for $\sim$23 ks and 
revolution 987 (2005 April, OBS 3) for $\sim$ 102 ks. 
The calibrated event lists for both the pn and the MOS observations 
were extracted 
using the EPCHAIN and EMCHAIN pipeline tasks provided by 
the \xmm Science Analysis System (SAS) version 9.0.0, using the 
latest calibration files available at the time of analysis
(2009 August). 
In our analysis we deal only with events corresponding to patterns 0-4 for 
the pn and 0-12 for the MOS instruments. 
We excluded the epochs of high background events with SAS task $espfilt$.
As an instance, we plot in Figure 1 the pn and MOS1 light curves of OBS1 (black), 
which was heavily polluted by high background flares, together with the derived good time intervals 
(green) by $espfilt$.   
The log of all the \xmm observations with net exposure times is shown in Table 1.
Details of both source and background spectra extraction are shown in Section 3.2.

\section{Results}
\label{fitting}

\subsection{Imaging and Radial Profiles} 
\label{prelimfit}

 In order to improve the signal-to-noise ratio (S/N) and fully utilize the spatial resolution 
of \xmm, { we 
extracted a co-added MOS image from three observations, which has better spatial resolution than pn detector,} 
and re-binned with a pixel size of 1.1$\arcsec$ (1 MOS pixel).
The resultant image is displayed in Figure 2 (a). 
The cross marks the optical center of the galaxy.
A bright point-like source (source X-1) is detected by \xmm 
about 25$\arcsec$\ away from the galaxy nucleus, which was originally detected in the $ROSAT$ All-Sky Survey, identified as a ULX by Colbert \& Ptak (2002). 

It can be seen from Figure 2 (a) that the emission of NGC 7590 is resolved at the 
\xmm spatial resolution, extending to tens of arcsec from the center. 
To quantify this, we first extract the radial profile of the X-ray emission from NGC 7590.  
To avoid the contamination from the bright off-nuclear X-1, only photons below the dashed 
line in Figure 2. (a) were extracted with 1.1\arcsec\ annular bins. The radial profile for X-1 was also extracted (without excluding the extended emission from NGC 7590) for comparison. The resulting radial profiles normalized to the peak brightness of NGC 7590 are shown in  Figure 2(c).
We then fitted the radial profiles with the
MOS point-spread function (PSF) model from calibration files using the SAS routine $eradial$.  
It is obvious in the figure that the radial profile of NGC 7590 significantly differs from that of  the off-nuclear source X-1 and MOS PSF model out to radii of $\sim$30 \arcsec, showing that the X-ray emission from NGC 7590 is clearly extended. 
The radial profile of the ULX is in crude agreement with the PSF model, as expected from a point source emission. 

We further used SciSim (Version 2.1) to simulate a MOS image of two point sources at the same position of the ULX and the nucleus of NGC 7590.
We first extracted the X-ray counts from the observed image for the 
ULX within an aperture 
responding to the $\sim$50\% encircled energy radius ($\sim$9$\arcsec$)\footnote{Which
 means $\sim$50\% of all the X-ray photons are encircled within this radius. See \xmm Users Handbook, Section 3.2.1}, and estimated 
 total net counts of $\sim1730$ for ULX after aperture correction. We then subtracted 1730 from the total counts of ULX plus NGC 7590 within 40$\arcsec$\ radius circle 
and obtained net counts of 
$\sim$ 1380 for NGC 7590 itself. In our simulation, we assumed that both NGC 7590 
and ULX would have a power law spectrum with $\Gamma=1.7$. Local background was 
added to the simulated image and the result is shown in Figure 2(b). By comparing 
the simulated image with the observed one, we see that the observed X-ray emission 
from NGC 7590 is clearly extended.

{ We also created a 
co-added pn image (excluding OBS 3)\footnote{There are bad columns across the galaxy nucleus in the pn image of OBS 3. We also excluded this exposure from the following spectral analysis 
for NGC 7590.}
with a pixel size of $4''$ (1 pn pixel), and extracted radial profiles from it for NGC 7590 and ULX. 
Figure 2(d) and (f) show the resulting pn image and radial profiles, 
which are similar to that of MOS, confirming that the X-ray emission from NGC 7590 is extended. 
We also present archival ROSAT HRI image for NGC 7590 (Figure 2(e)), which has even better 
spatial resolution than \xmm. However, the HRI image is too shallow 
for more detailed spatial analysis, and its soft X-ray coverage prevents us from estimating the nuclei hard X-ray emission.   
} 
  
{ There is possibility that the observed spatial extent of NGC 7590 and ULX is due to 
the alignment errors when adding images of three \xmm observations for both pn and MOS 
data respectively. 
To test this, we extracted the radial profile of X-ray emission of NGC 7590 and ULX, 
using only MOS image of OBS 3 (Figures 3.(a) and (d)), which has the longest exposure. 
Compared to Figures 2.(a) and (c), we do not found significant difference in the 
spatial extent of NGC 7590 and ULX, indicating it is unlikely that the observed extended 
emission of NGC 7590 is caused by image alignment errors. 
Figures. 3(b) and (c) show the soft-band (0.5--2 keV) and hard-band (2--10 keV) MOS 
images of OBS 3, and Figures (e) and (f) are the extracted radial profile, respectively. 
It can be seen that the spatial extent of soft-band emission of NGC 7590 and ULX 
agrees well with previous results. 
All these results suggest that the ULX is a point-like source, but the X-ray emission of 
NGC 7590 is clearly extended. 
Indeed, it is presumable that the hard-band image of NGC 7590 will be more point-like.  
However, as one can see from Figure. 3(c), the hard-band emission of NGC 7590 is 
relatively rather weak. Due to the poor S/N of the radial profile of NGC 7590 
(Figure. 3(f)), 
we cannot tell whether its hard-band emission is extended or more point-like. 

}

\subsection{Spectral Analysis}
$XMM$ spectra for NGC 7590 and X-1 were both extracted for spectral analysis.
For ULX X-1, to minimize the contamination from the extended emission in NGC 7590, 
we extracted source spectra using a 9$\arcsec$ radius circle (encircles 
$\sim$50\% of all the X-ray photons at off-axis angles less than 10 arcmin on 
the MOS detector, see Section 3.1) for both pn and MOS cameras, and aperture 
correction was applied after spectral fitting.
For NGC 7590, we extracted source spectra from a circular region with a 
radius of 40\arcsec\ centered at the galaxy nucleus, but excluded all photons 
above the dashed line in Figure 2(a) to reduce the pollution from X-1. 
Assuming that the X-ray emission from NGC 7590 is symmetric to the dashed line, 
hereafter we applied flux correction by a factor of 2 to the spectral fitting results to estimate the total X-ray flux from NGC 7590 (excluding X-1).
For both pn and MOS data, the background events were then extracted from a source-free area on the same CCD using two circular regions with a combined area $\sim$ 4 
times larger than the source region.
The response matrix and ancillary response file for the pn spectrum were 
generated using the RMFGEN and ARFGEN tools within the SAS software.

{ We examined the source light curves of NGC 7590 and ULX, and found no significant 
intra-exposure variations. No spectral variations between individual exposures were detected either through preliminary spectral fitting. 
Hereafter, we combined spectra from individual exposures to increase the spectral quality} 
and finally obtained one pn and one MOS spectra for each source (NGC 7590 and X-1).
The combined spectra were then grouped to have at least 1 count
 per bin, and the $C$-statistics (Cash 1979) was adopted for minimization.
Spectral fitting was performed jointly to the pn and MOS data in the 0.3--8 keV 
range using XSPEC (version 11.3.2). 
All statistical errors given hereafter correspond to 
90\% confidence for one interesting parameter ($\Delta \chi^2=2.706$), 
unless stated otherwise. In all of the model fitting, the 
Galactic column density was fixed at $N_{\rm H} = 1.96 \times
10^{20} \rm \ cm^{-2}$ (Dickey \& Lockman 1990). 
All model parameters will be referred to in the source frame.  

{\it NGC 7590:} 
We first performed the spectral fitting with a simple absorbed power law. 
The best-fitting parameters are $\Gamma=2.46^{+0.28}_{-0.23}$, and 
$N_{\rm H}<7\times10^{20}$ cm$^{-2}$, with $C$=520.8 for 436 degrees of freedom.
We note that the simple absorbed power law model does not provide a good 
representation of the data, leaving the significant  
residuals at the low energy of the spectra. X-ray studies of Seyfert
 2 galaxies have shown that their spectra often show a soft excess at lower energy band, 
presumably due to a thermal emission component (e.g., Cappi et al. 2006). Thus  
we added an optically-thin thermal emission component 
(using the APEC model in XSPEC) 
to the absorbed power law. A significantly improved fit 
($\Delta C=89$ for two extra parameters)
 is obtained with best-fitting temperature of the thermal component
 $kT$ = 0.31$^{+0.02}_{-0.03}$ keV. 
We then obtained the photon index $\Gamma=1.54^{+0.26}_{-0.27}$, while the
 column density is constrained to be less than $3\times10^{20}$ cm$^{-2}$.
When we add a narrow Gaussian line,   
with energy fixed at 6.4 keV (the rest energy of the \feka line) to the above model, 
the change of $\Delta C$ is 
less than 1 for addition of one free parameter, suggesting that 
the line is not detected.  
We therefore obtained an upper limit on the equivalent width (EW) of the line 
of $\sim$680 eV. 
The best-fitting model gives the observed 
$L_{0.5-2~\rm keV}=2.7\times10^{39}$ erg s$^{-1}$ and 
$L_{2-10~\rm keV}=2.9\times10^{39}$ 
erg s$^{-1}$. In Table 2, we show the best-fitting parameters, together with 
the model flux and luminosity in the 2-10 keV band.
\footnote {We note that Bassani et al. (1999) obtained a 
2--10 keV flux of 1.2 $\times$ 10$^{-12}$ \ergs for NGC 7590 based on \asca data. 
This is much larger than the sum of the \xmm flux of NGC 7590 and X-1, which is 
only $\sim$ 1.4 $\times$ 10$^{-13}$ \ergs. Through independent spectral fitting 
to archival \asca data, we obtained an \asca flux of $\sim$ 1.8 $\times$ 10$^{-13}$ \ergs, close to \xmm results. The flux presented by Bassani et al. for NGC 7590 is thus incorrect and could be a typo.}
 The spectra are shown in Figure 4 (upper panel), together with the best-fit model and the
 residuals from the best-fit model.

{ Spatial analysis (Section 3.1) has shown that the X-ray emission of NGC 7590 is dominated by both 
the ULX and an extended component from the host galaxy. 
To probe the true nuclear emission due to the AGN, we performed a spectral extraction
centered at the optically defined nucleus using a circular region of 9$\arcsec$ radius (50\% encircled-energy radius).
The derived spectra
could be best-fitted by an absorbed power law with N$_H<4\times10^{20}$ cm$^{-2}$ and
$\Gamma=1.66^{+0.35}_{-0.41}$, plus a thermal component $KT$$=0.32^{+0.10}_{-0.05}$ keV.
The resulting X-ray flux F$_{2-10~\rm keV}$ = 1.6$\times10^{-14}$ \ergs (after aperture correction), and the upper limit to Fe K$\alpha$ line EW rises to 2060 eV (see Table 2). 
Note that the derived flux could serve as an upper limit to the nuclear X-ray emission in NGC 7590.
}

{\it NGC~7590~X-1:} NGC 7590 X-1 is an extranuclear X-ray source, originally 
detected in the $ROSAT$ All-Sky Survey (Colbert \& Ptak 2002).
\xmm spectra are well fitted with an absorbed power-law model with 
$N_{\rm H}=5\pm2\times10^{21}$ cm$^{-2}$, $\Gamma=2.13^{+0.13}_{-0.12}$, 
and $C$/dof=657/758. The observed 0.5--2 keV and 2--10 keV fluxes (after aperture correction)
were $5.1\times10^{-14}$ \ergs and $9.3\times10^{-14}$ \ergs, respectively. 
If the source is associated with NGC 7590, the implied 2--10 keV luminosity is 
5.7$\times10^{39}$ \erg, making it as a ULX.  
The addition of a multicolor disk blackbody component 
is not statistically significant ($C$ decreased by only 2.6 for the addition of two free parameters). 
The \xmm spectra together with the best-fit model and residual are shown in Figure 4 (lower panel). 



\section{Discussion and Conclusions}
\label{conclusions}

NGC~7590 was previously identified as an "unobscured" Seyfert 2 galaxy, 
on the basis of the \asca spectrum (Bassani et al. 1999). 
However, the poor X-ray spatial resolution of \asca ($\sim$1$\arcmin$ 
PSF FWHM) casts doubt that the X-ray flux may be contaminated by nearby bright sources. 
Thanks to its better spatial resolution ($\sim$6$\arcsec$ PSF FWHM), the \xmm data  
clearly show that the X-ray emission from NGC 7590 is dominated by an off-nuclear ULX and an extended component from the host galaxy.

{ 
By performing a spectral extraction using a circular region of 9$\arcsec$ radius (50\% encircled-energy radius), we gave an upper limit to the nuclei X-ray emission F$_{2-10~\rm keV}$ = 1.6$\times10^{-14}$ \ergs\ (after aperture correction). 
Although the best fitting $N_{\rm H}$ is less than $4\times10^{20}$ cm$^{-2}$, we find  
an upper limit to the $T$ ratio ($f_{2-10~\rm keV}/f_{\rm O[ III]}$ $<$ 0.09)\footnote
{$f_{\rm O[ III]}$ has been extinction-corrected (Panessa \& Bassani 2002).}.
This value is in the range for the Compton-thick AGNs 
(Bassani et al. 1999; Guainazzi et al 2005).
The Fe K$\alpha$ line was not detected with \xmm and we could only obtain an upper limit of EW $<$ 2 keV, which is consistent with $\sim$1 keV 
expected in the case of a Compton-thick AGN. }

With \asca data, Nicastro, Martocchia \& Matt (2003) estimated that 
the accretion rate for NGC 7590 is very low ($<10^{-4}$), suggesting it may intrinsically 
lack a BLR. Meanwhile, NGC 7590 
 has not shown a hidden BLR in polarized light (Heisler et al. 1997). 
Tran (2001) finds that the Seyfert 2 galaxies without hidden 
BLRs are those with the lowest luminosity, while Shu et al. (2007) suggested 
that the non-detection of hidden BLRs is ascribed to the large obscuration.  
Making use of $L_{\rm X}$ measured by \xmm,
we attempted to derive the accretion rate for NGC 7590 to determine whether it is still consistent with the Nicastro et al. (2003) results.
 If NGC 7590 is explained as Compton thick, the direct measurement of its  
intrinsic X-ray luminosity seems to be impossible. 
Thus, with a correction factor of 60 to its intrinsic luminosity (e.g., 
Panessa et al. 2006) and a black hole mass of 
6.2$\times10^6~\rm M{\sun}$ derived from $M_{\rm BH}-\sigma_{\ast}$ 
relation (Bian \& Gu 2007), we obtain a rough estimate for NGC 7590 the accretion rate 
$\sim2\times10^{-3}$, a value close to the threshold ($\sim10^{-3}$) for 
BLR formation proposed by Nicastro et al. (2003). 
Note that Bian \& Gu (2007) suggested the Nicastro et al. Eddington ratio 
threshold corresponds to an $L_{\rm bol}/L_{\rm Edd}\sim10^{-1.37}$, if using 
[O {\sc iii}]$\lambda$5007 (rather than X-ray) luminosity to estimate the bolometric luminosity.  
The $L_{\rm bol}/L_{\rm Edd}$ of $10^{-1.4}$ found for NGC 7590 using [O {\sc iii}] 
luminosity is more in agreement with the Nicastro et al. (2003) predictions, suggesting that NGC 7590 is likely genuinely lack of a BLR. 
However, XMM images suggest a Compton-thick obscuration to the nucleus of NGC 7590, thus 
supports the scenario that the non-detection of hidden BLR in this source could be 
attributed to heavy obscuration (see Shu et al. 2007), and the 
intrinsic lack of BLR is not 
necessarily required.

The \xmm spectral fitting of NGC 7590 requires an additional soft component, 
which was well modeled with an emission spectrum 
from collisionally-ionized diffuse gas (the APEC model in XSPEC).  
The best-fit temperature for the soft thermal component is $\sim0.3$ keV. 
This value is consistent with $KT\sim0.2-0.8$ keV found in nearby 
Seyfert galaxies (Cappi et al. 2006), but lower than the range of gas temperatures 
(0.6--0.8 keV) found in LINERs (Gonz$\acute{a}$lez-Martin et al. 2006).   
The inferred soft X-ray luminosity is 
$L_{0.5-2~\rm keV}\sim2.7\times10^{39}$ \erg. 
If the soft X-ray emission is originated from gas heated by  
intense star formation activities,
the implied star formation rate for NGC 7590 is $\sim$ 1$M_{\sun}$ yr$^{-1}$ 
(Rovilos et al. 2009), which is consistent with the estimate using 
 its far-infrared luminosity (Kennicutt 1998). 

In summary, \xmm observations reveal that the X-ray emission of NGC 7590 is dominated 
by an extended component and an off-nuclear ULX. This indicates that its 
pre-identified "unobscured" nature is likely artificial. 
Due to the contamination from the ULX and the extended component, we are unable to 
isolate the nuclear X-ray emission for NGC 7590, and the flux upper limit was given.  
The derived relatively small $T$ ratio suggests that the galaxy likely has a heavily 
obscured nucleus, which could be confirmed with higher spatial resolution $Chandra$ images.
{ Future hard X-ray imaging telescopes (i.e., NUSTAR, ASTRO-H) could also help by detecting
the hard X-ray bump at $\sim$ 20 -- 30 keV of the Compton-thick spectrum.}


This research made use of the HEASARC online data archive services, supported
by NASA/GSFC.
The work was supported by Chinese NSF through Grant 10773010/10825312, and the 
Knowledge Innovation Program of CAS (Grant No. KJCX2-YW-T05).

\clearpage
\begin{figure}[!htb]
\epsscale{1.0}
\plotone{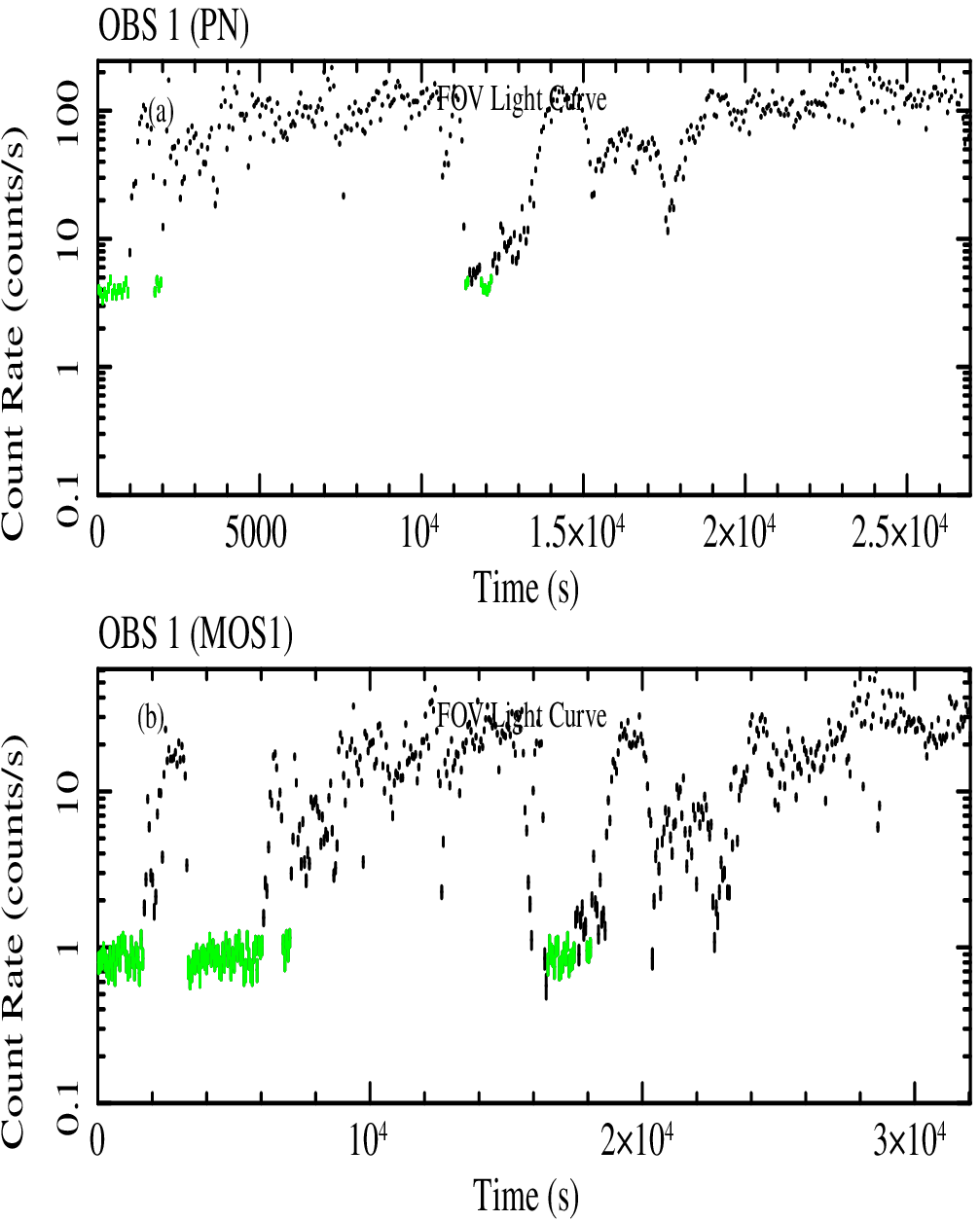}\caption{FOV light curves for OBS1 (black) and filtered good time intervals (green) by $espfilt$.
 }
\end{figure}

\begin{figure}[!htb]
\epsscale{1.0}
\plotone{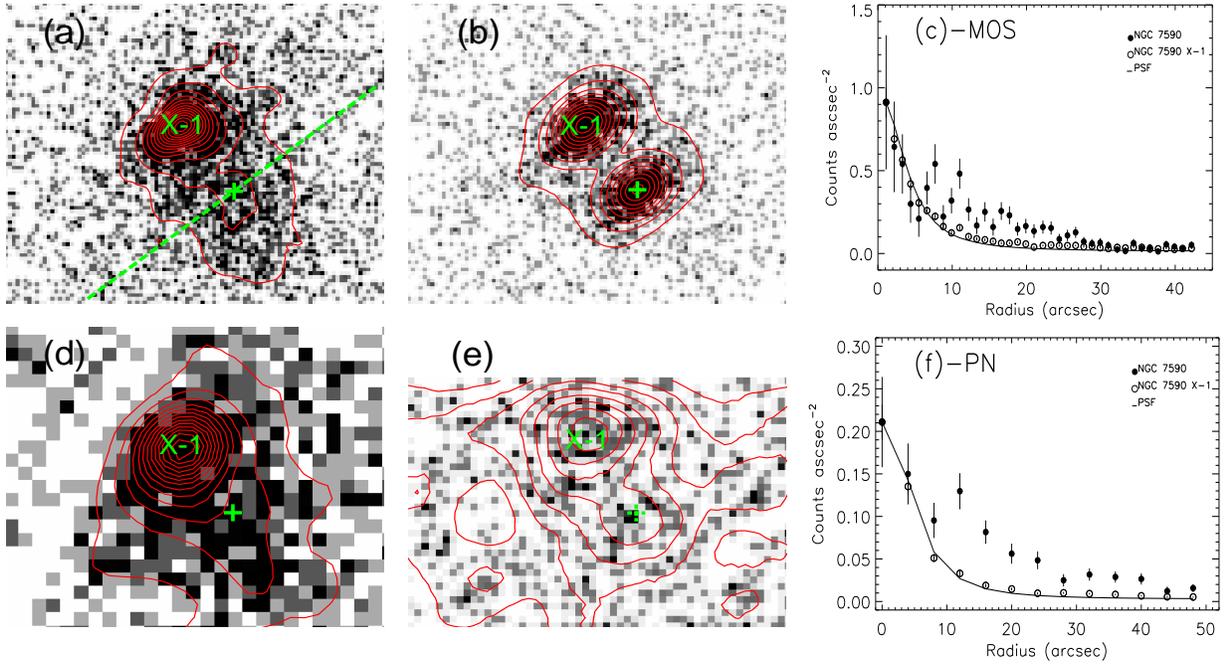}\caption{
{\it Left:} \xmm MOS (upper) and pn (lower) images of NGC 7590 with X-ray contours overlaid.
The cross denotes the optical nucleus of the galaxy. Source X-1 denotes the off-nuclear
ULX. {\it Middle:} Simulated EPIC MOS image of two point sources (upper). The lower panel is 
ROSAT HRI image.
{\it Right:} The observed MOS and pn radial profiles of the NGC 7590 (filled circles),
ULX (open circles), compared with the \xmm PSF (solid lines).The data points are normalized to the peak brightness of NGC 7590. While extracting radial profiles for NGC 7590, only photons below the dashed line in panel (a) were included to avoid contaminations from the ULX.}
\end{figure}

\begin{figure}[!htb]
\epsscale{1.0}
\plotone{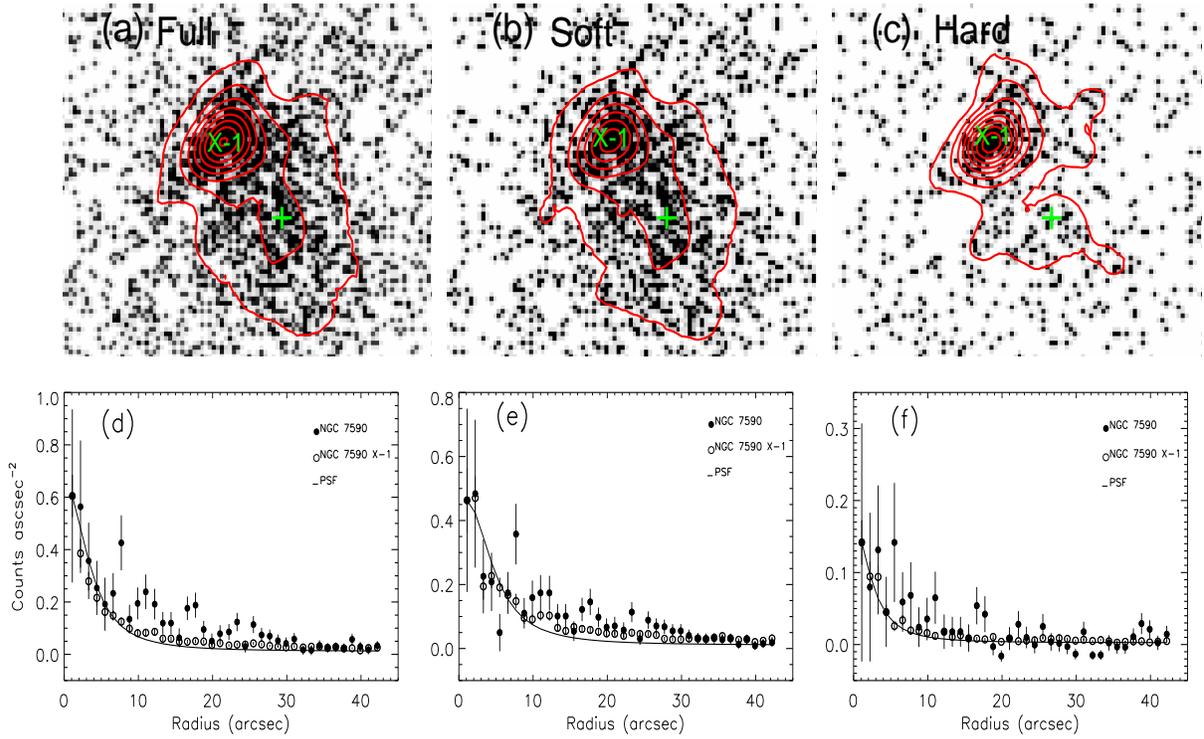}\caption{
Co-added MOS images for OBS 3 at different bands with X-ray contours overlaid (upper), and associated radial profile of 
ULX and NGC 7590, compared to the XMM PSF model (lower).   
The cross denotes the optical nucleus of the galaxy. Source X-1 denotes the off-nuclear
ULX. {\it Left:} 0.3-10 keV; {\it Middle:} 0.3-2 keV; {\it Right:}  
2-10 keV.
}
\end{figure}

\newpage

\begin{figure}[htb]
\epsscale{0.6}
\plotone{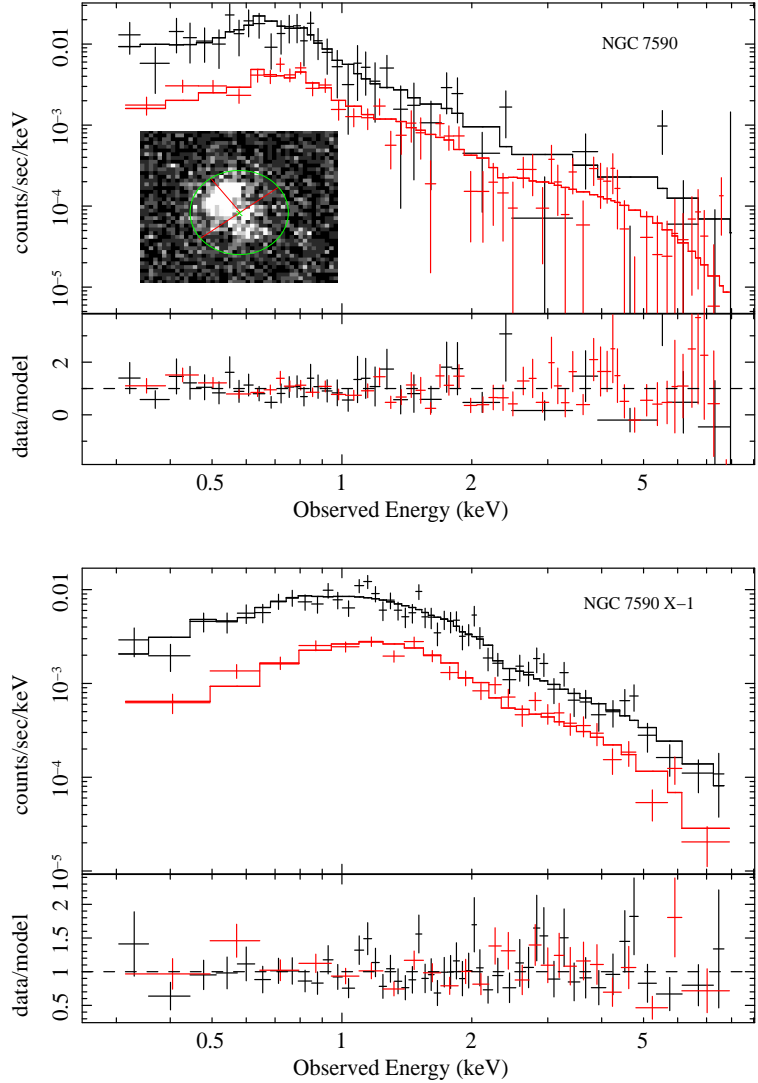}
\caption{EPIC pn (black) and MOS (red) spectra of NGC 7590 (upper) and ULX (lower),
together with the best-fitting model and the ratio of the data to the model.
The inset is the co-added pn and MOS image for OBS 2 to demonstrate
how the source spectra were extracted for NGC 7590 (see the text for details). }\end{figure}

\newpage
\clearpage
\begin{deluxetable}{lccr}
\tablecaption{\label{tab:lbgs}\xmm Observations of NGC 7590}
\tablewidth{0pt}
\tablehead{\multicolumn{1}{c}{ObsID} & \multicolumn{1}{c}{Rev.}
& \multicolumn{1}{c}{Date} &\multicolumn{1}{c}{Net. Exp. (ks)}  \\
   &  &  & (pn/MOS1/MOS2) }
\startdata
 0405380701 (OBS 1)& 1353 &   2007 Apr 30 & 1.5/5.8/5.8   \\
  0112310201 (OBS 2) & 0267 &   2001 May 25 & 13.4/20.2/21.1  \\
  0204610101 (OBS 3)& 0987 &   2005 Apr 29 & 52.2/64.5/65.1 \\
\enddata
\end{deluxetable}
\begin{deluxetable}{lllllllll}
\tablecaption{\label{tab:lbgs}Results of X-ray spectral fitting}\tablewidth{0pt}
\tablehead{\multicolumn{1}{c}{Target} & \multicolumn{1}{c}{$N_{\rm H}$}
 &\multicolumn{1}{c}{$\Gamma$} & \multicolumn{1}{c}{KT}
& \multicolumn{1}{c}{EW$_{6.4}$} &\multicolumn{1}{c}{F$_{2-10~keV}$} & \multicolumn{1}{c}{L$_{2
-10~keV}$}
&\multicolumn{1}{c}{$C/dof$} \\
 \hspace*{10.mm}(1) &  (2) & (3) &  (4) &  (5) & (6) & (7) & (8)  }
\startdata
ULX       &0.25$\pm$0.04 & 2.13$^{+0.13}_{-0.12}$ & $\dots$ & $\dots$ &
9.3 &5.7 & 657/758 \\
NGC 7590  & $<0.03$ & 1.54$^{+0.26}_{-0.27}$ & 0.31$^{+0.02}_{-0.03}$ &
  $<$680 & 4.8 & 2.9 & 431/433 \\
NGC 7590 (nucleus)  & $<0.04$ & 1.66$^{+0.35}_{-0.41}$ & 0.32$^{+0.10}_{-0.05}$ &
  $<$2060 & 1.6 & 1.0 & 185/168 \\

\enddata
\tablecomments{Col.(1): target.
Col.(2): column density of absorbed power-law in units of
10$^{22}$ cm$^{-2}$. Col.(3)
photon index. Col.(4): temperature of the thermal component in keV.
Col.(5): EW of \feka line in eV. Col.(6):
model flux in the 2--10 keV band, in units of 10$^{-14}$ \ergs (after aperture correction).
Col. (7): observed luminosity in the 2--10 keV
band, in units of 10$^{39}$\erg. Col. (8): $C$ and degrees of freedom.}
\end{deluxetable}

\end{document}